\documentclass[english]{article}
\usepackage[latin9]{inputenc}
\usepackage{color}
\usepackage{graphicx}
\usepackage{setspace}

\makeatletter

\DeclareRobustCommand{\greektext}{%
  \fontencoding{LGR}\selectfont\def\encodingdefault{LGR}}
\DeclareRobustCommand{\textgreek}[1]{\leavevmode{\greektext #1}}
\DeclareFontEncoding{LGR}{}{}
\DeclareTextSymbol{\~}{LGR}{126}

\makeatother

\usepackage{babel}
\begin{document}

\title{\textcolor{black}{Testing Quantised Inertia on Galactic Scales.}}

\author{\textcolor{black}{M.E. McCulloch}%
\thanks{\textcolor{black}{SMSE, Plymouth University, PL4 8AA, UK. mike.mcculloch@plymouth.ac.uk.}%
}}
\maketitle
\begin{abstract}
\begin{singlespace}
\noindent \textcolor{black}{Galaxies and galaxy clusters have rotational
velocities (v) apparently too fast to allow them to be gravitationally
bound by their visible matter (M). This has been attributed to the
presence of invisible (dark) matter, but so far this has not been
directly detected. Here, it is shown that a new model that modifies
inertial mass by assuming it is caused by Unruh radiation, which is
subject to a Hubble-scale ($\Theta$) Casimir effect predicts the
rotational velocity to be: $v^{4}=2GMc^{2}/\Theta$ (the Tully-Fisher
relation) where G is the gravitational constant, M is the baryonic
mass and c is the speed of light. The model predicts the outer rotational
velocity of dwarf and disk galaxies, and galaxy clusters, within error
bars, without dark matter or adjustable parameters, and makes a prediction
that local accelerations should remain above $2c^{2}/\Theta$ at a
galaxy's edge.}\end{singlespace}

\end{abstract}

\section{\textcolor{black}{Introduction}}

\begin{singlespace}
\noindent \textcolor{black}{Zwicky (1933) first noticed that galaxies
in galaxy clusters were moving too fast to be held together gravitationally
by their visible matter, and proposed the existence of an invisible
(dark) matter that provides the extra required gravitational pull.
A similar problem in disc galaxy rotation was proven by the accurate
rotation curves of Rubin }\textit{\textcolor{black}{et al}}\textcolor{black}{.
(1980). Dark matter is still the most popular explanation for these
problems, but, after decades of searching, it has not been directly
detected, though many efforts are ongoing, such as CDMS-II (2009)
and XENON10 (2009).}

\noindent \textcolor{black}{Milgrom (1983) proposed an alternative
explanation for galaxy rotation. He speculated that either 1) the
force of gravity may increase or 2) the inertial mass ($m_{i}$) may
decrease for the low accelerations at a galaxy's edge. His empirical
scheme, called Modified Newtonian Dynamics (MoND), can fit disc galaxy
rotation curves, and has the advantage of being less tunable than
dark matter. However, it does require one arbitrary parameter, the
acceleration $a_{0}$, and it does not predict the dynamics of galaxy
clusters (Aguirre et al., 2001, Sanders, 2002). The model proposed
in this paper has no adjustable parameters, and fits these clusters
more closely (it fits spirals less well, but is still within error
bars).}
\end{singlespace}

\noindent \textcolor{black}{Haisch et al. (1994) proposed that inertia
might be due to a reaction to the magnetic component of Unruh radiation,
which is seen only by accelerating bodies (the work of Haisch et al.
was an initial inspiration in the development of the model of inertia
presented in this paper, although the actual mechanism that produces
inertia is not specified here, and need not be that of Haisch et al.).}

\noindent \textcolor{black}{The wavelength of Unruh radiation lengthens
as acceleration reduces, and Milgrom (1994) noted that as galactic
radius increases, the rotational acceleration reduces, the Unruh waves
lengthen, and, at the radius where galactic dynamics start to become
non-Newtonian, the Unruh waves reach the Hubble scale. He speculated,
without assigning a specific cause, that this event may abruptly reduce
inertia, affecting dynamics and perhaps explaining MoND. However,
an abrupt loss of inertia at a particular galactic radius is not what
is seen. The observations show a more gradual deviation from Newtonian
behaviour.}

\begin{singlespace}
\noindent \textcolor{black}{McCulloch (2007) proposed a model for
inertia that could be called a Modification of inertia resulting from
a Hubble-scale Casimir effect (MiHsC) or Quantised Inertia. MiHsC
assumes that the inertial mass of an object is caused by Unruh radiation
resulting from its acceleration with respect to surrounding matter,
and that this radiation is subject to a Hubble-scale Casimir effect.
This means that only Unruh waves that fit exactly into twice the Hubble
diameter are allowed, so that an increasingly greater proportion of
the Unruh waves are disallowed as accelerations decrease and these
waves get longer, leading to a new gradual loss of inertia as acceleration
reduces. This loss of inertia is far more gradual than Milgrom's proposal,
discussed above. In MiHsC the inertial mass becomes}

\noindent \textcolor{black}{
\begin{equation}
m_{I}=m_{g}\left(1-\frac{\beta\pi^{2}c^{2}}{|a|\Theta}\right)\sim m_{g}\left(1-\frac{2c^{2}}{|a|\Theta}\right)
\end{equation}
}

\noindent \textcolor{black}{where $m_{g}$ is the gravitational mass,
$\beta=0.2$ (part of Wien's displacement law), c is the speed of
light, and $\Theta$ is the Hubble diameter ($2.7\times10^{26}m$,
from Freedman, 2001). For the derivation of Eq. 1 see McCulloch (2007)
and for a justification for the use of the modulus of the acceleration
see McCulloch (2008) and McCulloch (2011). MiHsC has now been tested
quite successfully on several anomalies that have been observed in
environments where accelerations are small (see McCulloch, 2007, 2008,
2010, 2011). MiHsC violates the equivalance principle, but not in
a way that could have been detected in a torsion balance experiment
(McCulloch, 2011).}

\noindent \textcolor{black}{McGaugh et al. (2009) studied the baryonic
mass of disc galaxies and showed that there were none with a baryonic
mass of less than $10^{9}M_{\odot}$. This minimum mass is also predicted
by MiHsC (McCulloch, 2010) since in MiHsC mutual accelerations must
always be above $2c^{2}/\Theta$ (close to the acceleration attributed
to dark energy). Using this prediction of a minimum acceleration,
in this paper MiHsC is applied to the rotation of a wider range of
cosmic structures.}
\end{singlespace}

\section{\textcolor{black}{Method and Results}}

\textcolor{black}{Starting with Newton's second law, and his gravity
law for a star of mass m orbiting in a galaxy of mass M}

\textcolor{black}{
\begin{equation}
F=m_{i}a=\frac{GMm}{r^{2}}
\end{equation}
}

\noindent \textcolor{black}{where $m_{i}$ is the inertial mass, $a$
is the rotational acceleration ($v^{2}/r$) of the star towards the
galactic centre, G is the gravity constant, and M is the mass within
a radius r, and m is the gravitational mass of a star. Replacing the
inertial mass following McCulloch (2007, 2008) (see Eq. 1, above)}

\textcolor{black}{
\begin{equation}
m\left(1-\frac{2c^{2}}{|A|\Theta}\right)a=\frac{GMm}{r^{2}}
\end{equation}
}

\noindent \textcolor{black}{where \textbar{}A\textbar{} is the total
of all the accelerations of the star relative to nearby matter. This
acceleration A can be seperated into a mean part (a) that is constant
in time, and a variable part (a') that is allowed to vary due to local
accelerations. Therefore
\begin{equation}
\left(1-\frac{2c^{2}}{(|a|+|a'|)\Theta}\right)a=\frac{GM}{r^{2}}
\end{equation}
}

\noindent \textcolor{black}{Multiplying through by $|a|+|a'|$ gives}

\noindent \textcolor{black}{
\begin{equation}
\left(|a|+|a'|-\frac{2c^{2}}{\Theta}\right)a=\frac{GM(|a|+|a'|)}{r^{2}}
\end{equation}
}

\noindent \textcolor{black}{MiHsC predicts a minimum allowed acceleration
of $2c^{2}/\Theta$. This occurs because as acceleration reduces,
the wavelength of the Unruh radiation increases and a greater proportion
of the waves are disallowed by the Hubble-scale Casimir effect in
MiHsC, so the inertia decreases and it becomes easier for the object
to be accelerated again by the same external force. A balance is predicted
to occur at a minimum acceleration of $2c^{2}/\Theta$ (see McCulloch,
2007, 2010). At a galaxy's edge the rotational acceleration $a$ is
smaller than this, so it is assumed here that this residual minimum
acceleration is found in smaller scale motions of the stars: in $a'$,
so at the galaxy's edge $a'=2c^{2}/\Theta$ and terms 2 and 3 on the
left hand side of Eq. 5 cancel, leaving}

\noindent \textcolor{black}{
\begin{equation}
a^{2}=\frac{GM(|a|+|a'|)}{r^{2}}
\end{equation}
}

\noindent \textcolor{black}{At the galaxy's edge the radius is large,
therefore the rotational acceleration is tiny, so $a\ll a'$ and so}

\textcolor{black}{
\begin{equation}
a^{2}=\frac{GM|a'|}{r^{2}}
\end{equation}
}

\noindent \textcolor{black}{Since $a=v^{2}/r$}

\textcolor{black}{
\begin{equation}
v^{4}=GM|a'|
\end{equation}
}

\noindent \textcolor{black}{Therefore MiHsC predicts a Tully-Fisher
relation (Tully \& Fisher, 1977) with a constant of $a'=2c^{2}/\Theta=6.7\pm0.6\times10^{-10}m/s^{2}$
(the uncertainty of 0.6 arises from uncertainties in the Hubble constant
of 9\%, taken from Freedman, 2001), so}

\textcolor{black}{
\begin{equation}
v^{4}=\frac{2GMc^{2}}{\Theta}
\end{equation}
}

\noindent \textcolor{black}{Figure 1 shows the observed rotation velocity
of galaxies and galaxy clusters binned into different baryonic mass
ranges as compiled by McGaugh }\textit{\textcolor{black}{et al}}\textcolor{black}{.
(2009). The x axis is the baryonic mass (gas plus stellar) in Solar
masses. The y axis (and the solid line with black circles) shows the
observed rotation (circular) speed (Vc) in km/s with an error of 30\%
caused by extrapolating Vc to larger radii (McGaugh }\textit{\textcolor{black}{et
al}}\textcolor{black}{., 2009).}

\noindent \textcolor{black}{The dashed line shows the predictions
of MiHsC from Eq. 9 which has a 20\% error because of: 1) a 9\% uncertainty
in the Hubble constant and therefore the value of $\Theta$ used,
and 2) a factor of 2 uncertainty in the stellar mass to light ratio
(McGaugh }\textit{\textcolor{black}{et al}}\textcolor{black}{., 2009)
which has a lesser impact for the darker low mass dwarf systems.}

\noindent \textcolor{black}{For dwarf galaxies (the three cases on
the left) the velocity predicted by MiHsC is higher than that observed
by up to 44\%, but MiHsC agrees with the observations, given the 30\%
uncertainty in the observed velocity and the 20\% uncertainty in the
prediction. For the spiral galaxies the predicted velocity is higher
than that observed by between 30-54\%. This is also in agreement given
the errors. For the galaxy clusters, MiHsC overpredicts the velocity
by 3-38\% but is still in agreement with the observations.}

\noindent \textcolor{black}{To summarise: MiHsC, which forbids accelerations
below $2c^{2}/\Theta$, can predict the rotation velocity of dwarf
galaxies, spiral galaxies and galaxy clusters within error bars, from
the baryonic mass only, and with no adjustable parameters (see Eq.
9), but tends to overpredict the velocities by between a third and
a half.}

\section{\textcolor{black}{Discussion}}

\textcolor{black}{The prediction of MoND is shown in Fig. 1 with the
two dotted lines, using the formula $v^{4}=\sqrt{GMa_{0}}$, and using
two values of $a_{0}$ to represent the range of values used in MoND:
$a_{0}=1.2\times10^{-10}m/s^{2}$ (the darker dotted line) and $a_{0}=2\times10^{-10}m/s^{2}$
(the lighter dotted line). MoND underestimates galaxy cluster velocities
and dark matter must be added to the clusters to fit MoND to the observations
(Sanders, 2002). If we assume error bars of 30\% on the observations
then both MoND and MiHsC agree with the data. MoND performs better
for spirals and MiHsC performs better for galaxy clusters. However,
MoND requires an unexplained adjustable parameter $a_{0}$ to fit
it to the data. With MiHsC no adjustable parameters are needed (see
Eq. 9).}

\textcolor{black}{When the mean acceleration $a$ is large (close
to the galactic centre) then the assumptions used here are not valid
and the prediction of MiHsC is Newtonian. As the galactic radius increases
the mean acceleration ($v^{2}/r$) decreases, but MiHsC predicts that
the acceleration cannot fall below $6.7\times10^{-10}m/s^{2}$ so
a testable prediction of MiHsC is that in the outer edges of a galaxy
smaller scale accelerations must increase to offset the decrease in
rotational accelerations with radius and keep accelerations above
$6.7\times10^{-10}m/s^{2}$.}

\textcolor{black}{This analysis assumes a balance between gravity
and the inertial force so is only valid for rotationally-supported
disc galaxies, and not elliptical galaxies and the galaxy clusters
(although MiHsC seems to predict those well). A similar derivation
could be tried for pressure-supported systems.}

\section{\textcolor{black}{Conclusion}}

\textcolor{black}{A relation between the velocity dispersion of a
disc galaxy or galaxy cluster and its visible mass (a Tully-Fisher
relation) can be derived by assuming that inertia is due to Unruh
radiation which is subject to a Hubble-scale Casimir effect (MiHsC).}

\textcolor{black}{The derived relation is, $v^{4}=2GMc^{2}/\Theta$,
and without adjustable parameters or dark matter, MiHsC predicts the
dispersion velocity of dwarf galaxies, spiral galaxies and galaxy
clusters within the error bars in these values (overestimating the
observed velocities typically by one third to one half) and predicts
that local accelerations should remain above $2c^{2}/\Theta$ at a
galaxy's edge.}

\section*{\textcolor{black}{Acknowledgements}}

\textcolor{black}{Thanks to S. McGaugh, K. Rosser and an anonymous
reviewer for advice, and B. Kim for encouragement.}

\section*{\textcolor{black}{References}}

\noindent \textcolor{black}{Aguirre, A., J. Schaye and E. Quataert,
2001. Problems for MoND in clusters and the LY\textgreek{a} forest?
}\textit{\textcolor{black}{Astrophys. J.}}\textcolor{black}{, 561,
550.}

\noindent \textcolor{black}{CDMS Collaboration, (Ahmed Z. }\textit{\textcolor{black}{et
al}}\textcolor{black}{.), }\textit{\textcolor{black}{Phys. Rev. Lett.}}\textcolor{black}{,
102, 011301.}

\begin{singlespace}
\noindent \textcolor{black}{Freedman, W.L., 2001. Final results of
the Hubble space telescope key project to measure the Hubble constant.
}\textit{\textcolor{black}{ApJ}}\textcolor{black}{, 553, 47-72.}

\noindent \textcolor{black}{Haisch, B., A.Rueda, H.E.Puthoff, 1994.
Inertia as a zero-point-field Lorentz force. }\textit{\textcolor{black}{Phys.
Rev. A}}\textcolor{black}{., 49, 678-694.}

\noindent \textcolor{black}{McCulloch,~M.E.,~2007. Modelling the
Pioneer anomaly as modified inertia. }\textit{\textcolor{black}{MNRAS}}\textcolor{black}{,
376, 338-342.}

\noindent \textcolor{black}{McCulloch, M.E., 2008. Modelling the flyby
anomalies using a modification of inertia. }\textit{\textcolor{black}{MNRAS-letters}}\textcolor{black}{,
389(1), L57-60.}

\noindent \textcolor{black}{McCulloch, M.E., 2010. Minimum accelerations
from quantised inertia. }\textit{\textcolor{black}{EPL}}\textcolor{black}{,
90, 29001.}
\end{singlespace}

\textcolor{black}{McCulloch, M.E., 2011. The Tajmar effect from quantised
inertia. }\textit{\textcolor{black}{EPL}}\textcolor{black}{, 95, 39002.}

\textcolor{black}{McGaugh, S.S., J.M. Schombert, W.J.G e Blok and
M.J. Zagursky, 2009. }\textit{\textcolor{black}{Astrophys. J.}}\textcolor{black}{,
708, L14.}

\begin{singlespace}
\noindent \textcolor{black}{Milgrom~M.,~1983. A modification of
the Newtonian dynamics as a possible alternative to the hidden mass
hypothesis. $Astrophysical~Journal$, 270, 365.}

\noindent \textcolor{black}{Milgrom, M., 1994. }\textit{\textcolor{black}{Ann.
Phys.}}\textcolor{black}{, 229, 384.}
\end{singlespace}

\noindent \textcolor{black}{Rubin, V., N. Thonnard, W.K. Ford Jr,
1980. Rotational properties of 21 Sc Galaxies with a large ange of
luminosities and radii from NGC 4605 (R=4kpc) to UGC 2885 (R=122kpc).
}\textit{\textcolor{black}{Astrophysical Journal}}\textcolor{black}{,
238, 471.}

\textcolor{black}{Sanders, R.H., 2002. Clusters of galaxies with Modified
Newtonian Dynamics (MoND). }\textit{\textcolor{black}{MNRAS}}\textcolor{black}{,
342, 3, 901-908.}

\textcolor{black}{Tully, R.B and J.R. Fisher, 1977. A new method of
determining distances to galaxies. }\textit{\textcolor{black}{Astronomy
and Astrophysics}}\textcolor{black}{, 54, 3, 661-673.}

\textcolor{black}{XENON10 Collaboration, }\textit{\textcolor{black}{Phys.
Rev. D.}}\textcolor{black}{, 80, 115005.}

\begin{singlespace}
\noindent \textcolor{black}{Zwicky,~F., 1933. Der Rotverschiebung
von extragalaktischen Nebeln. }\textit{\textcolor{black}{Helv. Phys.
Acta}}\textcolor{black}{, 6, 110.}
\end{singlespace}

\section*{\textcolor{black}{Figures}}

\textcolor{black}{\includegraphics[scale=0.72]{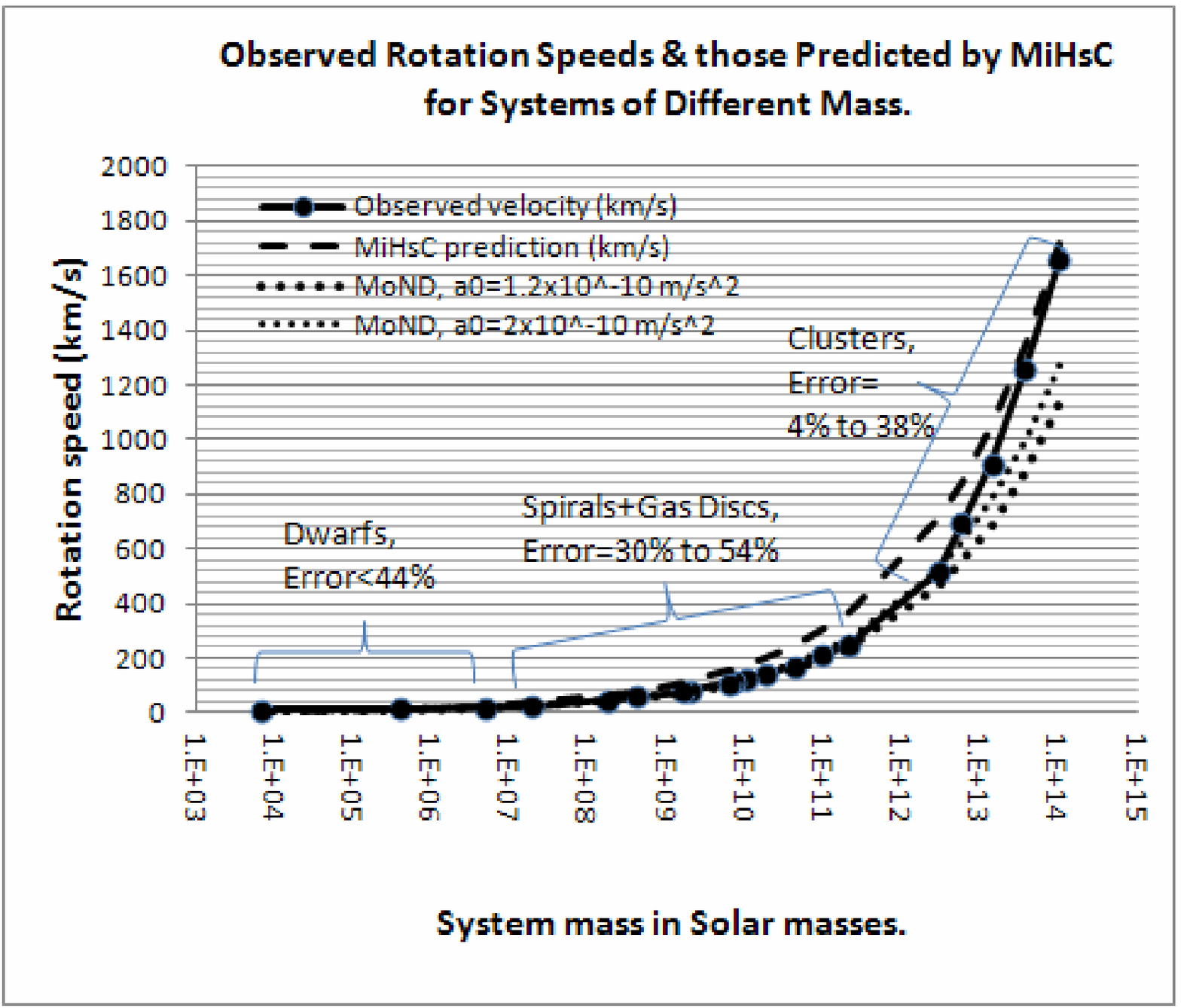}}

\textcolor{black}{Figure 1. The observed outer rotational velocity
for bins of galaxies and galaxy clusters of various masses from McGaugh
}\textit{\textcolor{black}{et al}}\textcolor{black}{. (2009) (black
circles, solid line) and the prediction of MoND (dotted line) and
MiHsC (dashed line). MiHsC overestimates the velocity for spirals
by 30-54\%, but outperforms MoND for galaxy clusters.}
\end{document}